\documentclass{aa}    

\usepackage{graphicx}
\usepackage{txfonts}

\usepackage{natbib}
\usepackage{dcolumn,url,longtable}
\usepackage{amssymb}
\usepackage{color}

\newcolumntype{d}{D{.}{.}{-1}}

\begin{document} 

  \title{Reconstruction of asteroid spin states from Gaia DR2 photometry}

  \author{J. \v{D}urech         \inst{1}        \and
          J. Hanu\v{s}          \inst{1}        
         }

  \institute{Astronomical Institute, Faculty of Mathematics and Physics, Charles University, V Hole\v{s}ovi\v{c}k\'ach 2, 180\,00 Prague 8, Czech Republic\\
             \email{durech@sirrah.troja.mff.cuni.cz}
             }

  \date{Received ?; accepted ?}

  \abstract
  % context heading (optional)
  {In addition to  stellar data, Gaia Data Release 2 (DR2) also contains  accurate astrometry and photometry of about 14,000 asteroids covering 22 months of observations.}
  % aims heading (mandatory)
  {We used Gaia asteroid photometry to reconstruct rotation periods, spin axis directions, and the coarse shapes of a subset of asteroids with enough observations. One of our aims was to test the reliability of the models with respect to the number of data points and to check the consistency of these models with independent data. Another aim was to produce new asteroid models to enlarge the sample of asteroids with known spin and shape.}
  % methods heading (mandatory)
  {We used the lightcurve inversion method to scan the period and pole parameter space to create final shape models that best reproduce the observed data. To search for the sidereal rotation period, we also used  a simpler model of a geometrically scattering triaxial ellipsoid.}
  % results heading (mandatory)
  {By processing about 5400 asteroids with at least ten observations in DR2, we derived models for 173 asteroids,  129 of which are new. Models of the remaining asteroids were already known from the inversion of independent data, and we used them for verification and error estimation. We also compared the formally best rotation periods based on Gaia data with those derived from dense lightcurves.}
  % conclusions heading (optional)
  {We show that a correct rotation period can be determined even when the number of observations $N$ is less than 20, but the rate of false solutions is high. For $N > 30$, the solution of the inverse problem is often successful and the parameters are likely to be correct in most cases. These results are very promising because the final Gaia catalogue should contain photometry for hundreds of thousands of asteroids,  typically with several tens of data points per object, which should be sufficient for reliable spin reconstruction.}

  \keywords{Minor planets, asteroids: general, Methods: data analysis, Techniques: photometric}

  \maketitle

  \section{Introduction}

  The ESA Gaia mission \citep{Gaia:16} has been in the science operations phase since August 2014. So far, there have been two Data Releases, the first  in September 2016 \citep[DR1,][]{Gaia:16b} and the second in April 2018 \citep[DR2,][]{Gaia:18}. The main output of DR2 is accurate astrometric data for more than a billion stars. However, unlike DR1, DR2 also contains astrometric and photometric data for about 14,000 asteroids \citep{Spo.ea:18}.

  Time-resolved photometry of asteroids,  i.e. lightcurves,  can be used for the reconstruction of the rotation period, spin axis orientation, and shape \citep[][for example]{Kaa.ea:02, Dur.ea:07b, Han.ea:16, Mar.ea:07, Mar.ea:18, Mic.ea:04, Kry:13}. Also, photometry that is sparse in time with respect to the rotation period can be successfully used with the same lightcurve inversion method \citep{Kaa:04, Dur.ea:07, Dur.ea:09, Dur.ea:16, Han.ea:11, Han.ea:13b}. Gaia provides this type of sparse-in-time photometry with unprecedented accuracy. After the end of mission, these data will be used to  determine periods, spins, and triaxial shape models \citep{Cel.ea:06, Cel.ea:07, Cel.Del:12, San.ea:15}. As shown by \cite{San.ea:15}, the probability of deriving a correct spin model is related to the shape (spherical asteroids have small lightcurve amplitudes), spin axis latitude (low-latitude asteroids are sometimes seen pole-on with small lightcurve amplitude), and the number of data points. Until now, real Gaia asteroid photometry was not available and the performance of inversion techniques was tested on  simulated data. DR2 has changed this situation and we can now use real high-quality Gaia photometry and test whether the expectations were met. Here we use asteroid photometry released in DR2 with the aim of testing the limits of lightcurve inversion and the information content of the data. We also derive new asteroid models.

  \section{Inversion of Gaia asteroid photometry}
  \label{sec:method}
    
    The DR2 contains G-band brightness measurements with uncertainties for about 14,000 asteroids. The observations cover 22 months and the number of data points per object varies from a few to 50. As described by \cite{Spo.ea:18}, the reported brightness values are constant for a single transit and they were computed as average values over the transit. \cite{Spo.ea:18} also tested  the accuracy of the  asteroid photometry  and reached  the conclusion that it is probably better than 1--2\%. This is much better than the accuracy of sparse photometry from ground-based surveys, which is hardly better than 0.1\,mag \citep{Dur.ea:09}. A unique reconstruction of the shape/spin model is possible only if there are enough photometric data points with good accuracy covering a sufficiently wide interval of geometries. With ground-based surveys, the poor photometric quality is compensated with the number of data points, typically several hundred, observed over many apparitions. Even so, unique solutions are rare;  the success rate of deriving a robust and reliable model is less than one percent \citep{Dur.ea:16}. In the case of Gaia, the final catalogue will contain data that fulfil all three requirements: they will be very accurate, there will be  several tens of measurements per object, and they will cover several apparitions for a typical main-belt asteroid. According to simulations, several tens of accurate measurements should be sufficient to derive a unique spin solution and an approximate shape \citep{San.ea:15}.

      \begin{figure*}[t]
        \begin{center}
          \includegraphics[height=0.47\textwidth]{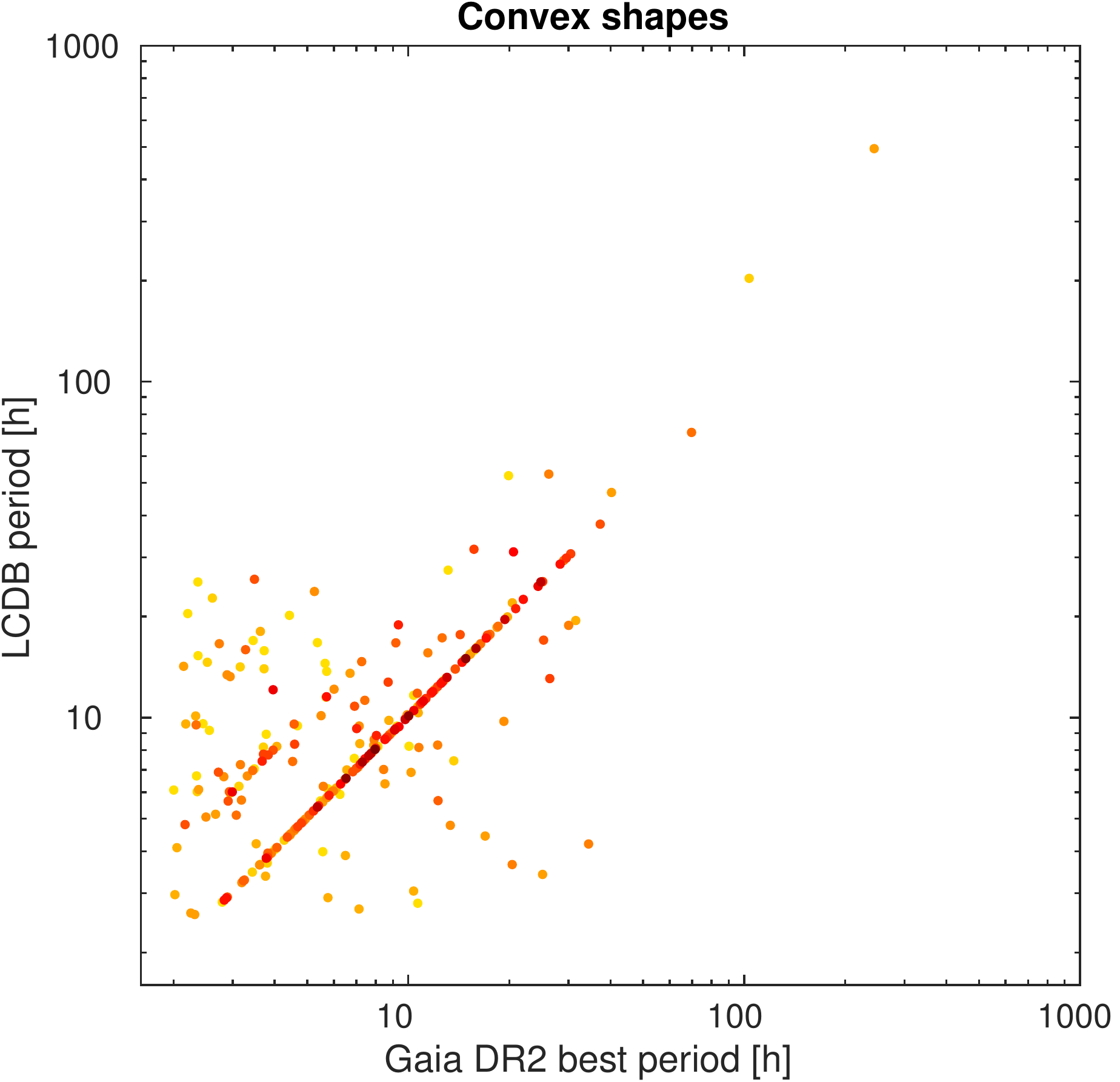}
          \includegraphics[height=0.47\textwidth]{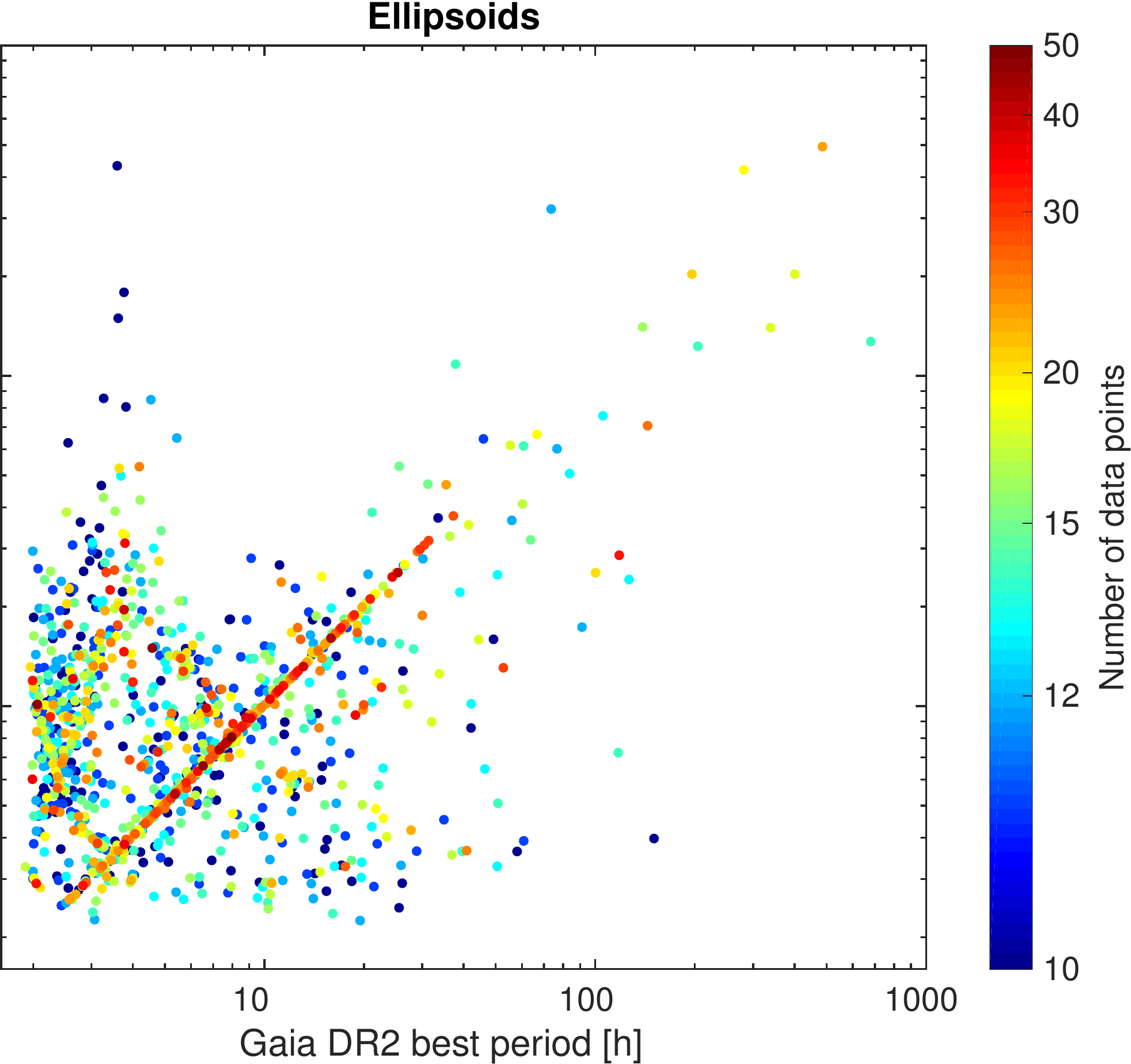}
          \caption{\label{fig:period_comparison} Comparison between the best period derived from Gaia photometry with convex models (left) and ellipsoids (right). Each point represents an asteroid for which the best-fitting period from Gaia was determined and with a period in the LCDB with the uncertainty code $\mathrm{U} \geq 3$.  The number of Gaia observations $N$ is expressed by the colour. The left panel contains fewer points because convex models were used only when $N > 20$.}
        \end{center}
      \end{figure*}

    DR2 photometry allowed us to test how successful the inversion of real Gaia data is. We selected all 5413 asteroids for which the number of brightness measurements $N$ was $\geq 10$. We computed the geometry with respect to the Sun and the Gaia spacecraft for each observation and processed the data in the same way as in \cite{Han.ea:11} and \cite{Dur.ea:16, Dur.ea:18c} by using the lightcurve inversion method of \cite{Kaa.ea:01}. Our approach is similar to that of \cite{Tor.ea:18} with the main difference that we do not deal with any error analysis. For each processed asteroid, we searched for a shape/spin model that gives the best fit to the data, measured by the lowest $\chi^2$ between the observed and modelled brightness. All data points were given the same weight; we did not take into account errors of individual measurements. The reason was that the relative formal errors are mostly below 2\% (90\% of all data points), and in this range the difference between, for example, 1\% and 0.1\% accuracy plays no role because the errors introduced by the model are larger (simplified shape approximation and scattering model assuming uniform albedo). 
    
    \subsection{Rotation periods}
    \label{sec:periods}

      As the first step, we computed periodograms using convex shapes and ellipsoids and tested the reliability of  the formally best-fit period. In Fig.~\ref{fig:period_comparison}, we show the comparison between the best period derived from DR2 using either convex shapes or ellipsoids and the values compiled in the Lightcurve Database (LCDB) of \cite{War.ea:09}; we used the  version from November 12, 2017. We used only reliable LCDB periods with the uncertainty code $\mathrm{U} \geq 3$. The colour-coding correlates with the number of data points $N$. Convex models do not provide any periods when $N \leq 20$ (see the discussion below). The points concentrating on the diagonal line represent the correctly determined periods (Gaia and LCDB periods are the same). The points off the diagonal are likely incorrect Gaia periods because LCDB records with $\mathrm{U} \geq 3$ should be reliable. The minor diagonal in the left panel are false solutions with the derived period being half of the real one;  this can happen with convex shapes as they produce lightcurves with only one minimum/maximum per rotation. Ellipsoidal models do not have this disadvantage of producing false half periods, but the periods based on a small number of points are often wrong. In general, when there are more data points, it is more likely that the derived period will be correct.

\begin{figure}
     \begin{center}
      \resizebox{\hsize}{!}{\includegraphics{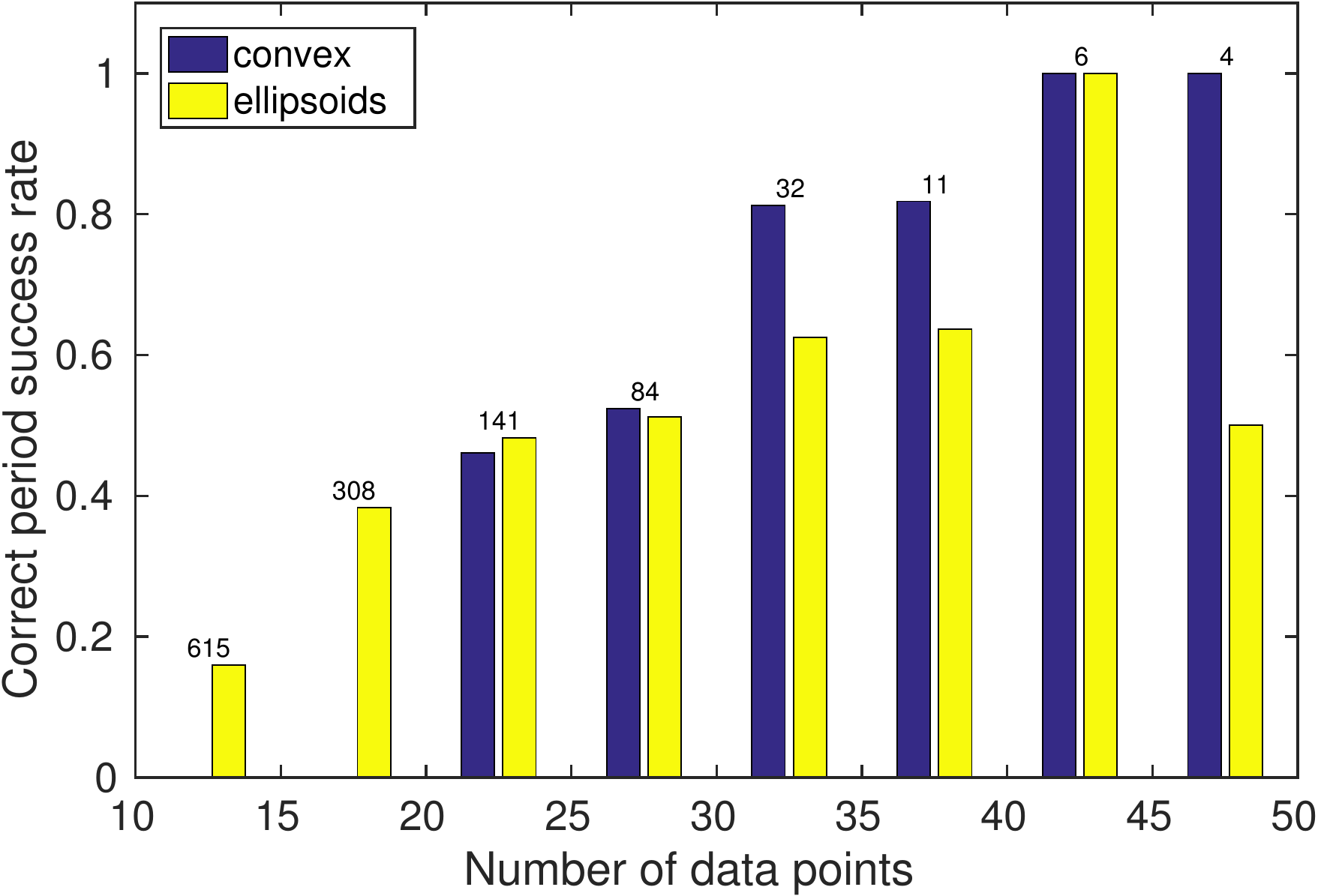}}
      \caption{\label{fig:period_success_rate} Success rate of deriving correct rotation periods (compared with LCDB) for ellipsoidal and convex models. The number above each histogram bar is the number of asteroids with periods in the LCDB with $\mathrm{U} \geq 3$ in that interval of data points.}
     \end{center}
    \end{figure}

      The dependence of the number of false periods on the number of data points is shown in Fig.~\ref{fig:period_success_rate}. The fraction of correctly determined periods (defined as those that agree with LCDB values within $\pm 5\%$) increases above 0.5 when $N > 30$. For fewer points, the formally best periods are not reliable. The clustering of points around shorter periods is likely a consequence of the way the model is constructed;  it is easier to formally fit the sparse points with an incorrect period that is shorter than the true period. For $N > 40$, the success rate seems to be high, but the sample size of asteroids in this range is very small.

      We also tested if there is any difference in the photometric errors of Gaia data between the asteroids with correctly and incorrectly determined rotation periods. For the two bins with $N$ between 21--25 and 26--30 (where the fraction of correctly determined periods is about 50\% and the number of periods is large), we compared the photometric errors of points belonging to asteroids with correctly determined periods with those that belong to asteroids with incorrect Gaia-based periods. The t-test did not reveal any significant difference in the means of these two groups. Also, the distribution of observations in time was very similar for the two groups. We did not reveal any statistical difference in, for example, the number of observations separated by $\sim 100$ minutes, which is the spacing related to the scanning pattern of Gaia corresponding to two field-of-view transits.

    \subsection{Spins and shapes}
    
      The best-fit periods discussed above are often just random global minima in the periodograms. To distinguish between random and real periods, we have to define some level of significance measured by the $\chi^2$ fit. We defined the uniqueness of the best solution by the depth of the $\chi^2$ minimum with respect to other local minima. The formula we used for the threshold $\chi^2_\text{tr} = (1 + \sqrt{2/\nu})\,\chi^2_\text{min}$ is a modification of the formula we used in \cite{Dur.ea:18c};  now there is no factor of 1/2. This is an arbitrary borderline that is based on a trade-off between the total number of new models and their contamination with incorrect models. Here $\nu$ is the number of degrees of freedom, which is formally the difference between the number of points $N$ and the number of parameters $p$. For ellipsoids, $p = 6$ and the parameters are the sidereal rotation period $P$, the spin axis direction in ecliptic coordinates $\lambda$ and $\beta$, one parameter for the linear slope of the phase curve, and two parameters ($a/c$ and $b/c$) for the ellipsoid axes ratios. With Gaia observations the phase angle is almost always $> 10\deg$, which means that  the phase function can be reduced to only a linear part with one parameter \citep{Kaa.ea:01}. With convex models, we used the spherical harmonics representation of the order  and degree of three \citep{Kaa.ea:01}, which corresponds to 16 shape parameters, so the total number of parameters is $p = 20$. For spin/shape reconstruction, we only used  asteroids with $N > 20$.

      When the number of points was small, in many cases we obtained RMS residuals of almost zero for many different periods. Such periodograms were excluded from the analysis. To avoid fitting noise, we only selected  periodograms with all RMS values $> 0.005\,$mag. Another requirement was that there should be only one period with RMS below 0.01. If there were more, we considered it  a non-unique solution even if the threshold limit was satisfied. The verification procedure was the same as in \cite{Dur.ea:18c}, see Fig.~3 there, with the only difference that we did not use $n = 6$ for the degree and order of the spherical harmonics series. The visual inspection of the periodograms was crucial because in many cases we obtained false solutions for $P \lesssim 3$\,h or $P \gtrsim 100$\,h. Even with a correct rotation period and pole, the corresponding shapes were often unrealistic with sharp edges and triangular pole-on silhouettes. This is a consequence of the  order and degree of the spherical harmonics series ($n = 3$) being too low, but with a small number of photometric points there is not enough information to reconstruct higher resolution models. In this sense, convex shape models derived from DR2 data should not be taken as real shapes;  they are just models that fit the data  best with the given resolution, and they are likely to change significantly when more data points are available and a higher degree of resolution is possible. On the other hand, the rotation periods and pole directions are not that sensitive to the resolution and they are more reliable \citep{Han.Dur:12}.

    \subsection{Comparison with independent models}
    \label{sec:models_comparison}

      By processing all asteroids with $N > 20$ and rejecting unreliable solutions, we derived models of 173 asteroids. Of these, 44 were already in the Database of Asteroid Models from Inversion Techniques \citep[DAMIT,][]{Dur.ea:10} and we used them for an independent test of the accuracy of our solutions based on DR2. Most of our models agreed with those in DAMIT: their periods were the same within the errorbars and the mean difference between their pole directions was $20^\circ$. However, there was a group of clear outliers with differences between pole directions $> 60^\circ$. We looked in detail into these cases (seven in total). In one case -- asteroid (2802)~Weisell -- the DAMIT model based only on sparse data \citep{Han.ea:16} was clearly incorrect because the period search was done in the wrong local minimum. We removed this model from DAMIT. 
      In five cases, the periods were very similar but differed more than their uncertainty, so the DR2-based solution was apparently a different local minimum leading to a different pole. In one case, the periods were completely different. 

    \subsection{New models}
    \label{sec:new_models}

    In Table~\ref{tab:models}, we list 129 new models and their spin axis directions (sometimes there are two possible solutions), the sidereal rotation period, and the period reported in the LCDB. The LCDB period agrees with our value in most cases. The asteroids for which the periods do not agree and the LCDB period is reliable (higher uncertainty code U) are marked with an asterisk.
    
    To further check the reliability of these new models, we repeated the period search using reduced data sets. For each asteroid from Table~\ref{tab:models}, we randomly selected and removed 10\% of the data points (i.e. 2--5 points) from the original data set. In most cases, this reduction had no effect and the new periodogram showed the same unique period. In 44 cases, the reduced data set still provided the same best period, but it did not pass the threshold limit and thus was not considered  a unique solution. In one case the new period was different from the original one. These asteroids are marked with an exclamation mark. This test shows that with our definition of $\chi^2_\mathrm{tr}$, we are at the critical limit of the number of data points in many cases. Removing just a few points may lead to formal rejection of the model even if it is correct (as is independently confirmed by the agreement between $P$ and $P_\mathrm{LCDB}$). 
    
    Because the geometry of observations is restricted mostly to the ecliptic plane, the lightcurve inversion usually produces two mirror shape solutions with about the same pole latitude $\beta$ and the difference in longitude of $\sim180^\circ$ \citep{Kaa.Lam:06}. However, there are a surprising number of  solutions with just one pole direction in Table~\ref{tab:models} \citep[compared with the results of][for example]{Dur.ea:18c}. This is likely caused by loosely constrained shapes that often are too elongated along the rotation axis. These shapes are removed by the pipeline. It means that the mirror pole solutions cannot be rejected even if they are not listed in Table~\ref{tab:models}.

  \section{Discussion}
  \label{sec:discussion}

    Our analysis of Gaia DR2 asteroid photometry shows that the excellent photometric accuracy enables us to derive reliable spin directions and rotation periods from the first 22 months of Gaia observations. Although the number of unique models derived from DR2 is small compared to the number of all asteroids with photometry, this is mainly because the number of asteroids with $\gtrsim 30$ measurements is still limited. However, the prospect for the next data releases is very high. With more than 50 points covering several years, the inversion should provide unique results in most cases (apart from very spherical asteroids or those with extreme rotation) and the shape models will be more robust. Moreover, the photometric calibration in the future data releases should be even more accurate due to further improvement of the reduction pipeline that in the case of DR2 did not include correction for flux loss due to moving objects or a more sophisticated filtering of outliers, for example \citep{Spo.ea:18}.
    
    Before DR3 (likely in the first half of 2021), the full potential of DR2 can be exploited when Gaia photometry is combined with archived lightcurves or sparse photometry from ground-based surveys. In this way, even a small number of accurate Gaia photometric measurements with a higher statistical weight can help to reconstruct uniquely the shape and spin state of many asteroids with the lightcurve inversion method.

  \begin{acknowledgements}
    The authors were supported by the grant no. 18-04514J of the Czech Science Foundation. This work has made use of data from the European Space Agency (ESA) mission {\it Gaia} (\url{https://www.cosmos.esa.int/gaia}), processed by the {\it Gaia} Data Processing and Analysis Consortium (DPAC, \url{https://www.cosmos.esa.int/web/gaia/dpac/consortium}). Funding for the DPAC has been provided by national institutions, in particular the institutions participating in the {\it Gaia} Multilateral Agreement.

  \end{acknowledgements}

  \newcommand{\SortNoop}[1]{}

  \begin{appendix}
  \onecolumn
   
    \section{List of new models}

    \tiny
    \begin{longtable}{r l r r r r @{} d @{} d l r c}
      \caption{\label{tab:models} List of new asteroid models. For each asteroid, we list one or two pole directions in the ecliptic coordinates $(\lambda, \beta)$, the sidereal rotation period $P$, the rotation period from LCDB $P_\mathrm{LCDB}$ (if available) and its quality code $U$, the number $N$ of sparse photometric data points in DR2, and the method  used to derive the rotation period: C -- convex inversion, E -- ellipsoids, CE -- both methods gave the same unique period. The accuracy of the sidereal rotation period $P$ is of the order of the last decimal place given. For asteroids marked with an asterisk, there is an inconsistency between $P$ and $P_\mathrm{LCDB}$ and those marked with an exclamation mark did not pass the $\chi^2_\mathrm{tr}$ limit when 10\% of points were removed.}\\
      \hline\hline
      \multicolumn{2}{c}{Asteroid}              & \multicolumn{1}{c}{$\lambda_1$}       & \multicolumn{1}{c}{$\beta_1$}   & \multicolumn{1}{c}{$\lambda_2$}       & \multicolumn{1}{c}{$\beta_2$}   & \multicolumn{1}{c}{$P$}       & \multicolumn{1}{c}{$P_\mathrm{LCDB}$} & \multicolumn{1}{c}{U}   & \multicolumn{1}{c}{$N$}       & method        \\
      number    & name/designation              & \multicolumn{1}{c}{[deg]}             & \multicolumn{1}{c}{[deg]}       & \multicolumn{1}{c}{[deg]}             & \multicolumn{1}{c}{[deg]}       & \multicolumn{1}{c}{[h]}       & \multicolumn{1}{c}{[h]}                 &                       &                               &               \\
      \hline
      \endfirsthead
      \caption{continued.}\\
      \hline\hline
      \multicolumn{2}{c}{Asteroid}              & \multicolumn{1}{c}{$\lambda_1$}       & \multicolumn{1}{c}{$\beta_1$}   & \multicolumn{1}{c}{$\lambda_2$}       & \multicolumn{1}{c}{$\beta_2$}   & \multicolumn{1}{c}{$P$}       & \multicolumn{1}{c}{$P_\mathrm{LCDB}$} & \multicolumn{1}{c}{U}   & \multicolumn{1}{c}{$N$}       & method        \\
      number    & name/designation              & \multicolumn{1}{c}{[deg]}             & \multicolumn{1}{c}{[deg]}       & \multicolumn{1}{c}{[deg]}             & \multicolumn{1}{c}{[deg]}       & \multicolumn{1}{c}{[h]}       & \multicolumn{1}{c}{[h]}                 &                       &                               &               \\
      \hline
      \endhead
      \hline
      \endfoot
                    205 & Martha                    &   358 & $-35$ &       &       &  14.9117 &      14.911 &     3 &    46 &   C  \\
          !   217 & Eudora                    &   131 & $ -4$ &   317 &    11 &   25.262 &      25.272 &     3 &    29 &   C  \\
              333 & Badenia                   &     3 & $-48$ &   179 & $-27$ &   9.8609 &       9.862 &     3 &    25 &  CE  \\
              561 & Ingwelde                  &    89 &    68 &       &       &  12.0118 &      12.012 &     3 &    22 &   C  \\
              580 & Selene                    &    62 &    66 &   269 &    50 &   9.4929 &        9.47 &  3$-$ &    34 &   C  \\
              581 & Tauntonia                 &   203 & $-49$ &       &       &   24.994 &       16.54 &     2 &    32 &   C  \\
              659 & Nestor                    &     6 & $-83$ &       &       &   15.979 &       15.98 &     3 &    42 &   C  \\
        !     723 & Hammonia                  &   146 &    22 &   330 &    21 &   5.4348 &       5.436 &     3 &    31 &  CE  \\
$\ast$    !   838 & Seraphina                 &    18 &     4 &   192 &    32 &  11.7245 &       15.67 &     2 &    28 &   C  \\
        !     842 & Kerstin                   &    18 &    78 &       &       &   18.716 &             &       &    23 &  CE  \\
              876 & Scott                     &   105 & $-18$ &   262 & $-35$ &  11.8178 &      11.814 &     2 &    27 &   E  \\
              906 & Repsolda                  &   108 & $-48$ &       &       &   15.367 &      15.368 &     3 &    23 &   E  \\
              961 & Gunnie                    &    37 &    24 &   220 &     7 &   21.361 &             &       &    23 &  CE  \\
              976 & Benjamina                 &   354 &    80 &       &       &   9.7080 &       9.700 &  3$-$ &    23 &   E  \\
             1029 & La Plata                  &   119 &    30 &   274 &    44 &  15.3101 &      15.310 &     3 &    21 &   C  \\
        !    1107 & Lictoria                  &    82 &    57 &   303 &    57 &   8.5610 &      8.5616 &     3 &    31 &  CE  \\
             1118 & Hanskya                   &   224 & $-87$ &       &       &   25.305 &       15.61 &     2 &    39 &   C  \\
$\ast$    !  1165 & Imprinetta                &    39 & $-82$ &       &       &  10.8087 &       8.107 &     3 &    25 &   C  \\
        !    1168 & Brandia                   &   310 &    65 &       &       &  11.4425 &      11.444 &     3 &    21 &   E  \\
        !    1220 & Crocus                    &    26 &    48 &   165 &    80 &    488.6 &       491.4 &     3 &    23 &   E  \\
             1431 & Luanda                    &    79 &    54 &       &       &  4.13591 &       4.141 &  3$-$ &    23 &   E  \\
             1437 & Diomedes                  &   320 &     1 &       &       &   24.501 &       24.49 &  3$-$ &    26 &   E  \\
             1533 & Saimaa                    &   332 & $-70$ &       &       &   7.1174 &        7.08 &     3 &    25 &   C  \\
$\ast$  !    1540 & Kevola                    &    96 & $-71$ &       &       &  5.00453 &      20.082 &  3$-$ &    32 &   E  \\
             1542 & Schalen                   &    92 &    41 &   268 &    52 &   7.5153 &       7.516 &     3 &    23 &   E  \\
        !    1604 & Tombaugh                  &   166 &    34 &   323 &    36 &   7.0359 &       7.047 &    2+ &    35 &  CE  \\
        !    1647 & Menelaus                  &   157 &    26 &   330 &    23 &   17.745 &       17.74 &  3$-$ &    28 &   E  \\
          !  1762 & Russell                   &     8 &    89 &       &       &  12.7933 &      12.797 &  3$-$ &    22 &   C  \\
        !    1767 & Lampland                  &   186 & $-53$ &   340 & $-47$ &   35.399 &             &       &    29 &  CE  \\
$\ast$    !  1786 & Raahe                     &    88 &    45 &       &       &   30.175 &       18.72 &     3 &    25 &   C  \\
        !    1799 & Koussevitzky              &    58 & $-61$ &       &       &   6.3256 &       6.318 &     3 &    22 &   E  \\
             1849 & Kresak                    &   140 &    61 &       &       &   31.716 &      19.101 &     2 &    27 &   E  \\
             1873 & Agenor                    &   117 &    56 &       &       &   20.631 &       20.60 &     2 &    26 &   E  \\
          !  1939 & Loretta                   &    21 & $-72$ &   201 & $-72$ &   23.931 &         25. &     1 &    33 &  CE  \\
          !  1975 & Pikelner                  &    74 &    38 &       &       &  11.2768 &             &       &    32 &   C  \\
             2090 & Mizuho                    &   231 & $-22$ &       &       &   5.4793 &        5.47 &    2+ &    22 &   E  \\
             2104 & Toronto                   &   306 & $-74$ &       &       &   8.9667 &        8.97 &     3 &    24 &   E  \\
             2111 & Tselina                   &   102 &    19 &   282 &    53 &   6.5631 &       6.563 &     3 &    46 &   C  \\
          !  2127 & Tanya                     &    64 &    61 &       &       &   7.8523 &       7.864 &     2 &    35 &  CE  \\
             2147 & Kharadze                  &   180 & $-13$ &   347 & $-45$ &  14.1384 &      14.115 &     2 &    26 &  CE  \\
             2192 & Pyatigoriya               &   139 &    44 &   338 &    81 &   8.7162 &             &       &    27 &   C  \\
             2203 & van Rhijn                 &    63 & $-76$ &       &       &   30.607 &       30.55 &     2 &    21 &   E  \\
             2230 & Yunnan                    &     2 &    64 &   154 &    73 &  10.0381 &             &       &    21 &   E  \\
        !    2386 & Nikonov                   &    52 &    51 &   242 &    33 &   5.8944 &             &       &    21 &   E  \\
             2397 & Lappajarvi                &    77 & $-55$ &   251 & $-35$ &   9.0532 &        9.05 &     2 &    28 &   C  \\
             2429 & Schurer                   &   235 & $-26$ &       &       &   6.5119 &        6.66 &  3$-$ &    42 &  CE  \\
        !    2587 & Gardner                   &   351 &    49 &       &       &  11.6268 &      11.631 &     2 &    30 &   E  \\
             2627 & Churyumov                 &   141 & $-45$ &   307 & $-70$ &   7.6531 &        7.66 &  3$-$ &    29 &  CE  \\
             2634 & James Bradley             &   120 & $-58$ &       &       &   16.514 &             &       &    21 &   E  \\
             2683 & Brian                     &   112 & $-34$ &   294 & $-50$ &   22.528 &             &       &    33 &  CE  \\
             2686 & Linda Susan               &    51 & $-52$ &   268 & $-78$ &   8.7222 &             &       &    31 &   C  \\
$\ast$       2760 & Kacha                     &   101 & $-30$ &       &       &   53.040 &        13.0 &     3 &    29 &   E  \\
             2884 & Reddish                   &   201 & $-84$ &       &       &   35.537 &             &       &    26 &  CE  \\
             3131 & Mason-Dixon               &   294 &    57 &       &       &   19.703 &      19.748 &     2 &    41 &   C  \\
             3134 & Kostinsky                 &   276 &    88 &       &       &  14.6755 &        14.7 &     2 &    41 &  CE  \\
             3210 & Lupishko                  &    42 &    55 &       &       &  14.2490 &      14.241 &     2 &    35 &   E  \\
             3325 & TARDIS                    &   142 & $-50$ &   338 & $-79$ &  11.5681 &             &       &    48 &  CE  \\
             3374 & Namur                     &   217 &    34 &       &       &  13.8542 &             &       &    27 &   C  \\
        !    3420 & Standish                  &   355 &    76 &       &       &  10.3869 &             &       &    25 &  CE  \\
             3451 & Mentor                    &    81 &    18 &       &       &   7.6966 &       7.702 &     3 &    29 &   E  \\
          !  3525 & Paul                      &    10 & $ -3$ &   214 &    27 &  13.4116 &         12. &  2$-$ &    21 &   C  \\
             3565 & Ojima                     &    99 &    66 &   294 &    68 &  15.4510 &             &       &    32 &   C  \\
        !    3776 & Vartiovuori               &   190 & $-83$ &       &       &  11.3414 &         7.7 &  2$-$ &    24 &   E  \\
             3788 & Steyaert                  &    67 & $-86$ &       &       &   29.966 &             &       &    29 &  CE  \\
          !  4075 & Sviridov                  &    17 &    89 &   329 &    62 &   6.6626 &             &       &    22 &   C  \\
             4131 & Stasik                    &    31 & $-64$ &   203 & $-70$ &  16.5405 &             &       &    25 &  CE  \\
             4271 & Novosibirsk               &   105 &    58 &   320 &    60 &   8.8469 &       8.850 &     3 &    27 &  CE  \\
        !    4352 & Kyoto                     &   155 &    42 &   341 &    28 &   21.922 &     21.9352 &  3$-$ &    32 &  CE  \\
          !  4366 & Venikagan                 &    14 &    38 &   188 &    37 &    60.48 &             &       &    32 &   C  \\
             4369 & Seifert                   &    15 & $-30$ &   170 & $-56$ &   30.617 &      30.573 &     3 &    29 &   C  \\
        !    4451 & Grieve                    &   177 & $-68$ &   341 & $-63$ &   6.8600 &       6.864 &     3 &    28 &   E  \\
          !  4575 & Broman                    &   181 & $-66$ &   303 & $-61$ &  10.7735 &             &       &    26 &  CE  \\
             4613 & Mamoru                    &   280 & $-53$ &       &       &  5.38872 &       5.388 &     3 &    41 &  CE  \\
             4732 & Froeschle                 &   105 & $-25$ &   289 & $-24$ &  12.0479 &             &       &    26 &   E  \\
        !    4930 & Rephiltim                 &   264 & $-72$ &       &       &  5.24273 &       5.243 &     3 &    30 &  CE  \\
        !    5059 & Saroma                    &    33 & $-59$ &   215 & $-30$ &   4.0744 &       4.074 &     3 &    22 &   E  \\
             5130 & Ilioneus                  &   128 & $-19$ &   303 & $ -8$ &  14.7357 &      14.768 &     3 &    22 &   E  \\
             5138 & Gyoda                     &   157 & $-90$ &       &       &   8.4159 &             &       &    23 &   E  \\
        !    5285 & Krethon                   &   182 & $-89$ &       &       &  12.0247 &       12.04 &     2 &    24 &   E  \\
             5344 & Ryabov                    &     9 &    90 &   286 &    71 &   18.490 &             &       &    26 &   E  \\
        !    5385 & Kamenka                   &   168 &    23 &       &       &   6.6719 &       6.683 &     2 &    25 &   E  \\
             5594 & Jimmiller                 &   260 & $-31$ &       &       &   25.275 &      25.264 &     2 &    26 &   C  \\
             5755 & 1992 OP7                  &    51 &    78 &   190 &    88 &   5.5752 &             &       &    29 &   E  \\
             5883 & Josephblack               &    38 &     7 &   228 &    21 &   17.542 &             &       &    27 &  CE  \\
             6173 & Jimwestphal               &    93 & $-30$ &       &       &  2.90849 &       2.908 &     3 &    27 &   C  \\
             6338 & Isaosato                  &   146 &    84 &       &       &  4.97689 &             &       &    30 &  CE  \\
             6665 & Kagawa                    &    40 & $-36$ &       &       &   6.7494 &             &       &    24 &   E  \\
        !    6794 & Masuisakura               &   154 & $-29$ &   345 & $-57$ &  4.58811 &        4.58 &     3 &    23 &   E  \\
        !    7022 & 1992 JN4                  &     4 &    84 &   337 &    56 &   5.5154 &       5.517 &     2 &    28 &   E  \\
             7238 & Kobori                    &    42 & $-83$ &   238 & $-88$ &  17.2242 &             &       &    37 &   C  \\
             7457 & Veselov                   &    16 &    88 &       &       &  12.2002 &             &       &    25 &   E  \\
             7458 & 1984 DE1                  &    48 &    39 &   232 &    17 &  15.7543 &        16.7 &     2 &    25 &   C  \\
          !  7616 & Sadako                    &   185 & $-70$ &   339 & $-37$ &  11.0085 &             &       &    24 &   C  \\
        !    7650 & Kaname                    &    58 & $-86$ &       &       &   18.177 &      18.172 &     2 &    24 &  CE  \\
        !    8066 & Poldimeri                 &   205 & $-69$ &   342 & $-88$ &   18.420 &             &       &    28 &   E  \\
             8292 & 1992 SU14                 &    32 &    20 &       &       &   19.802 &       2.856 &     1 &    25 &   E  \\
             8443 & Svecica                   &   261 &    48 &       &       &   20.994 &      20.998 &     3 &    31 &   C  \\
             8770 & Totanus                   &   118 & $-67$ &   294 & $-69$ &   14.867 &             &       &    41 &   C  \\
             9299 & Vinceteri                 &   228 & $-63$ &       &       &    87.95 &             &       &    22 &   E  \\
            10406 & 1997 WZ29                 &    14 & $-86$ &   184 & $-86$ &   6.7520 &             &       &    22 &   E  \\
            10763 & Hlawka                    &   128 &    30 &       &       &   6.6833 &             &       &    21 &   E  \\
            10790 & 1991 XS                   &   171 &    52 &   357 &    55 &   8.4894 &             &       &    33 &   C  \\
        !   11429 & Demodokus                 &    31 & $-77$ &       &       &    50.23 &       50.16 &     2 &    25 &   E  \\
          ! 11682 & Shiwaku                   &    25 &    68 &   206 &    74 &  4.01885 &             &       &    21 &  CE  \\
            12003 & Hideosugai                &    32 & $-90$ &       &       &   9.0157 &             &       &    27 &   E  \\
            12291 & Gohnaumann                &   252 &    52 &       &       &  3.22070 &             &       &    22 &   C  \\
            13446 & Almarkim                  &    48 &    54 &   230 &    53 &   5.5116 &             &       &    25 &   E  \\
            13809 & 1998 XJ40                 &    66 & $-27$ &   225 & $-30$ &  3.81363 &      40.256 &     2 &    22 &   E  \\
            14268 & 2000 AK156                &   119 & $-11$ &       &       &   7.5121 &        7.51 &  3$-$ &    26 &   E  \\
          ! 14362 & 1988 MH                   &   188 &     6 &       &       &  3.64387 &       3.639 &     3 &    21 &   C  \\
            14376 & 1989 ST10                 &   139 &    58 &   340 &    68 &   5.8481 &       5.614 &     3 &    28 &   E  \\
        !   14410 & 1991 RR1                  &   342 & $-76$ &       &       &  12.4272 &             &       &    32 &   E  \\
            15105 & 2000 BJ4                  &   108 &    71 &       &       &   6.2570 &       6.257 &     2 &    29 &  CE  \\
            15496 & 1999 DQ3                  &    23 &    89 &       &       &    81.21 &             &       &    27 &   E  \\
            15955 & Johannesgmunden           &   327 & $-85$ &       &       &   9.4985 &             &       &    22 &   E  \\
            16029 & 1999 DQ6                  &    94 &    85 &       &       &   5.9537 &        5.95 &  3$-$ &    25 &   E  \\
        !   16771 & 1996 UQ3                  &    49 &    63 &   253 &    41 &  4.95609 &             &       &    28 &  CE  \\
            17567 & Hoshinoyakata             &   325 & $-87$ &       &       &   19.891 &      19.882 &     2 &    22 &   E  \\
            18156 & Kamisaibara               &    23 & $-88$ &   221 & $-55$ &  11.9166 &             &       &    24 &   C  \\
            18666 & 1998 FT53                 &   316 &    49 &       &       &   5.5970 &             &       &    36 &  CE  \\
            20721 & 1999 XA105                &    74 & $-33$ &   259 &    11 &   5.3478 &             &       &    34 &   C  \\
        !   21904 & 1999 VV12                 &   225 &    73 &       &       &   5.4324 &             &       &    22 &   E  \\
            22972 & 1999 VR12                 &    90 &    57 &       &       &   22.659 &             &       &    27 &  CE  \\
        !   24324 & 2000 AT51                 &   132 &    53 &       &       &   5.8951 &             &       &    26 &   E  \\
            25846 & 2000 EF93                 &   123 &    27 &   300 &    55 &  16.9713 &             &       &    29 &   E  \\
            32497 & 2000 XF18                 &   214 &    15 &       &       &   24.578 &             &       &    22 &   E  \\
            40165 & 1998 QP102                &   331 &    86 &       &       &   7.4192 &       7.419 &     2 &    28 &   E  \\
            47678 & 2000 CT75                 &   159 &    54 &       &       &   9.4505 &       9.453 &     2 &    23 &   E  \\
        !   51857 & 2001 OA105                &    78 &    59 &   257 &    29 &  4.31349 &       4.313 &     2 &    23 &  CE  \\

    \end{longtable}
  
  \end{appendix}
\end{document}